\documentclass{emulateapj}

\usepackage{subfigure,epsfig,url}

\begin{document}

\title{Tracking Neptune's Migration History through High-Perihelion Resonant Trans-Neptunian Objects}

\author{Nathan A. Kaib\altaffilmark{1} \& Scott S. Sheppard\altaffilmark{2}}

\altaffiltext{1}{HL Dodge Department of Physics \& Astronomy, University of Oklahoma, Norman, OK 73019, USA}
\altaffiltext{2}{Department of Terrestrial Magnetism, Carnegie Institution for Science, 5241 Broad Branch Road, NW, Washington, DC 20015, USA}

\begin{abstract}

Recently, \citet{shep16} presented the discovery of seven new trans-Neptunian objects with moderate eccentricities, perihelia beyond 40 AU, and semimajor axes beyond 50 AU. Like the few previously known objects on similar orbits, these objects' semimajor axes are just beyond the Kuiper belt edge and clustered around Neptunian mean motion resonances (MMRs). These objects likely obtained their observed orbits while trapped within MMRs, when the Kozai-Lidov mechanism raised their perihelia and weakened NeptuneÕs dynamical influence. Using numerical simulations that model the production of this population, we find that high-perihelion objects near Neptunian MMRs can constrain the nature and timescale of Neptune's past orbital migration. In particular, the population near the 3:1 MMR (near 62 AU) is especially useful due to its large population and short dynamical evolution timescale. If Neptune finishes migrating within $\sim$100 Myrs or less, we predict over $\sim$90\% of high-perihelion objects near the 3:1 MMR will have semimajor axes within 1 AU of each other, very near the modern resonance's center. On the other hand, if NeptuneÕs migration takes $\sim$300 Myrs, we expect $\sim$50\% of this population to reside in dynamically fossilized orbits over $\sim$1 AU closer to the Sun than the modern resonance. We highlight 2015 KH$_{162}$ as a likely member of this fossilized 3:1 population. Under any plausible migration scenario, nearly all high-perihelion objects in resonances beyond the 4:1 MMR (near 76 AU) reach their orbits well after Neptune stops migrating and comprise a recently generated, dynamically active population.

{\bf Keywords:} 

\end{abstract}

\section{Introduction}

As Neptune gravitationally scatters small planetesimals, it will on average move further from the Sun due to exchange of angular momentum with the planetesimals \citep{fernip84}. Thus, in the presence of a Kuiper belt, Neptune's semimajor axis must increase over time as it interacts with dynamically unstable Kuiper belt objects. Because the modern Kuiper belt's mass is so anemic \citep{glad01}, Neptune's current semimajor axis is effectively fixed. However, the early solar system is thought to have possessed a much more massive Kuiper belt \citep[e.g.][]{stern96}, and under these conditions Neptune's semimajor axis could have evolved dramatically.

\citet{mal93} provided support for this scenario by showing that an outward migration of Neptune by $\sim$5 AU could capture Pluto into the 3:2 resonance with Neptune and subsequently excite its eccentricity to Pluto's modern value. Furthermore, \citet{mal95} demonstrated that Neptune's migration through a massive primordial Kuiper belt should result in the capture of many objects into the Neptunian resonances. The prominence of resonant populations has since been confirmed with the observed orbital distributions of detected TNOs, further supporting the idea that Neptune underwent substantial orbital migration early in its history \citep[e.g.][]{ell05, glad12, bann16}. 

While these resonant populations were initially interpreted as evidence of Neptune's relatively smooth, steady migration through a dynamically cold primordial Kuiper belt, an alternate view known as the Nice Model emerged in which the early orbital evolution of the giant planets could have been much more violent \citep{tsig05}. In this scenario, planets can gravitationally scatter off one another before having their orbits recircularized via dynamical friction from the still massive primordial Kuiper belt \citep{thommes99, thommes02, bras09}. The formation of the modern Kuiper belt from a massive primordial belt has been modeled under this violent scenario for the giant planets' early evolution, and it was found to reproduce many of the main features of the modern Kuiper belt \citep{lev08}. However, during such a violent giant planet instability, Neptune can become quite eccentric ($e>0.3$) for an extended period, and this can excite the orbits of the presumably primordial cold classical belt objects orbiting between 42 and 47 AU with low eccentricities and inclinations \citep{bat11, dawclay12}. More recent refinements of the giant planet instability have uncovered scenarios in which Neptune's eccentricity stays at or below $e\sim0.1$, and these can potentially preserve the dynamical state of the cold classical population \citep{nesmorb12, dawclay12}. 

Thus, many different models of Neptune's early orbital evolution have been proposed, and the detailed structure of the modern Kuiper belt provides some of the most robust constraints on it. For instance, \citet{nes15b} showed that the Kuiper belt's inclination distribution rules out Neptune having migrated more than $\sim$6 AU through the massive, primordial Kuiper belt, or migrating with an $e$-folding timescale shorter than $\sim$10 Myrs. Furthermore, it has recently been argued that this migration was interrupted with a ``jump'' in Neptune's semimajor axis caused by a scattering event between Neptune and another similar mass ice giant. This jump would cause objects trapped in the 2:1 resonance with Neptune to be suddenly released, providing an explanation for the curious concentration of cold classical objects near $\sim$44 AU \citep{petit11}. Moreover, \citet{nes16} found additional evidence that the rest of Neptune's migration was not completely smooth, either. With perfectly smooth migration, Kuiper belt formation models predict many more objects in the 3:2 resonance than observed today \citep{hahnmal05, lev08, nes15b}. Instead, \citet{nes16} were able to reproduce the correct 3:2 population size if an element of ``graininess'' was included in Neptune's migration. This graininess would arise as encounters between Neptune and the thousands of Pluto-mass objects caused sudden, small changes in Neptune's semimajor axis as it migrated. In order to account for the existence of large TNOs like Pluto and Eris \citep{brown05} today, there were likely many more present in the primordial Kuiper belt.

While many earlier and contemporary works have focused on the capture of resonant populations closer to the Sun than the 2:1 MMR at $\sim$48 AU, objects can also be captured into more distant resonances during Kuiper belt formation \citep[e.g.][]{dunlev97, gomes05, gall06, lykmuk07}. Indeed objects occupying these more distant resonances also appear to be abundant \citep{pike15, volk16}. When trapped within resonances with Neptune, Kozai-Lidov cycles can be activated if an object's inclination exceeds a critical value \citep{gomes03, koz62, lid62}. During these Kozai cycles, an orbit's eccentricity and inclination oscillate exactly out of phase with one another as the longitude of perihelion librates \citep{koz62, gomes05}. During the high-inclination phase of these Kozai cycles, an object's perihelion is lifted away from Neptune as its eccentricity decreases and its semimajor axis remains constant, and it becomes more weakly coupled to the planet. If the semimajor axis of the MMR is large enough, this process can raise the perihelia of TNOs well beyond 40 AU \citep{gomes08}. During these high-perihelion excursions, objects can even temporarily leave their resonances (as measured by resonant angle libration) \citep{gomes05}. 

Because the classical Kuiper belt stops near 48 AU, these high-perihelion objects with $a\gtrsim50$ AU stand out. Ignoring the potential effects of distant perturbers on extreme Kuiper belt objects \citep[e.g.][]{fern97, trushep14, batbrown16}, an object with $a\gtrsim50$ AU must typically have a perihelion near Neptune for it to have been scattered to a large orbit. These resonant high-perihelion objects are an exception to this rule. Examples of such objects have been detected, and perhaps the most well-known is 2004 XR$_{190}$ with a semimajor axis of 57.6 AU, perihelion of 51.5 AU, and an inclination of 46.6$^{\circ}$ \citep{allen06}. 2004 XR$_{190}$ orbits just interior to the 8:3 resonance location today, and \citet{gomes11} demonstrated that this object could have gone through Kozai cycles within the 8:3 MMR while Neptune was still migrating only to become decoupled during one of its excursions to large perihelion. After this decoupling, the orbital elements of 2004 XR$_{190}$ would remain essentially frozen while the location of the 8:3 resonance continued to move outward with Neptune's migration, explaining this object's location relative to the modern 8:3 MMR as well as its high inclination and perihelion. Recent work by \citet{brasil14} has shown that such evolutionary paths are also possible near the 5:2 and 3:1 MMRs. 

Thus, Neptune's migration can generate a dynamically fossilized population of high-perihelion objects adjacent to modern resonance locations. As such, this class of resonant or nearly resonant high-perihelion objects may provide additional constraints on the nature of Neptune's migration. Until recently, only a few of these objects were known. However, recent survey results by \citet{shep16} have added another 6 objects, increasing the statistical significance of this sample. With this is mind, we numerically model the production of high-perihelion objects in or near mean motion resonances with Neptune in order to assess this population's utility as a constraint on Neptune's migration. In our numerical models we vary both the timescale and smoothness of Neptune's migration to see the effects on the population of high-perihelion objects near Neptunian MMRs. Our work is organized in the following manner: Section 2 details the numerical methods we use in our simulations. Section 3 presents the results of this numerical work. Finally, Section 4 summarizes our conclusions.

\section{Numerical Methods}

Our numerical models of the Kuiper belt's formation are very similar to the models explored in \citet{nes15a, nes15b} and \citet{nes16}. We model the dispersal of a primordial Kuiper belt by Neptune as it migrates from 24 to 30 AU, similar to what is envisioned in the late stages of some realizations of the Nice model \citep{nesmorb12}. We simulate Kuiper belt formation under four different Neptune migration scenarios, exploring both the speed and smoothness of Neptune's migration. Our first two simulations both assume that Neptune migrates smoothly but at different speeds; we have one ``slow'' simulation and one ``fast'' simulation. In these two simulations, Neptune smoothly migrates from 24 to 28 AU with a fixed migration $e$-folding timescale of $\tau_1$. At this point it ``jumps'' by $\sim$0.5 AU. Neptune's semimajor axis is instantaneously increased by 0.5 AU and its eccentricity is increased from $\sim$0 to 0.075. This is meant to simulate a scattering event between Neptune and another approximately Neptune-mass ice giant in the early solar system \citep{nes11, nesmorb12, nes15b}. Following this jump, Neptune then migrates the rest of the way to 30 AU with a migration $e$-folding time ($\tau_2$) that is $\sim$3 times longer than $\tau_1$. Once Neptune reaches its current semimajor axis of 30.11 AU, migration is shut off for the remainder of the simulation. Our second two simulations are essentially repeats of our slow and fast simulations, except that an element of granularity is now superimposed onto Neptune's migration by adding small, random, instantaneous shifts to Neptune's semimajor axis as it migrates with the prescription given above. This is meant to simulate the effects of scattering events between Neptune and the numerous Pluto-sized objects thought to have resided in the primordial Kuiper belt \citep{nes16}. 

To run each of our Kuiper belt formation simulations, we use the SWIFT RMVS4 integrator \citep{levdun94} to evolve 1 million test particles under the influence of the four giant planets for 4 Gyrs. Jupiter, Saturn, and Uranus are all started on their current semimajor axes with small eccentricities ($e  < 0.01$) and inclinations ($i < 0.1^{\circ}$). The initial nearly circular, coplanar configurations for the inner 3 giant planets are chosen because Neptune's subsequent migration causes a moderate ($e\sim0.03$) increase in Uranus' eccentricity during the course of a simulation. Although this results in low values for the inner giant planets' eccentricities and inclinations at late times, the Kuiper belt's structure is largely insensitive to these planets' eccentricities and inclinations \citep{nes15a}. Meanwhile, Neptune is begun with a similarly small eccentricity and inclination, but its initial semimajor axis is set to 24 AU. The 1 million test particles we employ are randomly distributed along a $a^{-1}$ surface density profile between 24 and 30 AU, while eccentricities are uniformly drawn between 0 and 0.01. Inclinations of test particles are drawn randomly from the following distribution function

\begin{equation}
f(i) = \sin{i}\exp{-\frac{i}{2 \sigma^2}}
\end{equation}
where $\sigma = 1^{\circ}$. All other orbital elements are drawn randomly from an isotropic distribution. The particles and planets are evolved for 4 Gyrs with a timestep of 200 days, and test particles are removed due to collisions with the Sun or planets as well as if they exceed a heliocentric distance of 1000 AU. 

To force Neptune's semimajor axis to migrate, we employ simple drag forces on its velocity as in \citet{mal95}. This causes Neptune's semimajor axis to evolve in time according to the following:

\begin{equation}
a(t) = a_{f} + (a_{0} - a_{f}) \times \exp\left({-\frac{t}{\tau}}\right)
\end{equation}
where $a_{f}$ is the desired final semimajor axis, $a_{0}$ is the initial semimajor axis, and $\tau$ is the desired e-folding timescale. In all of our simulations, $a_{f}$ was initially set to 30.11 AU and $a_{0}$ was initially set to 24 AU.

In addition to semimajor axis migration, Neptune's eccentricity and inclination are also damped by employing the fictitious forces described in \citet{kom05}. With these included, both inclination and eccentricity decay exponentially. The magnitude of these forces was adjusted so that our eccentricity and inclination $e$-folding timescales were approximately equal to the semimajor axis migration $e$-folding timescale employed in Equation 2. 

Once Neptune reaches a semimajor axis of 28 AU, each simulation is temporarily stopped and Neptune's semimajor axis is instantaneously increased by 0.5 AU and its eccentricity is increased to 0.075, while all other orbital elements are held fixed. Then the simulation is restarted with migration and damping timescales that are longer by a factor of $\sim$3. This is continued until Neptune reaches a semimajor axis of 30.1 AU. At this point, all migration and damping forces are shut off, and the simulation is continued for the rest of the integration time using the standard RMVS4 package.

In the case of our grainy migration simulations, we add in thousands of tiny discrete jumps into Neptune's semimajor axis evolution. This technique was first employed in \citet{nes16}. To determine the magnitude and frequency of semimajor axis jumps, \citet{nes16} carefully studied individual encounters between Neptune and Pluto-mass objects as the planet migrated through a disk of 1000+ of these bodies. From this, a distribution of semimajor axis jumps was constructed as well as encounter times. Then, these distributions of encounter times and semimajor axis jumps were used to generate small, instantaneous orbital jumps for Neptune as it migrated through $\sim$10$^6$ test particles in RMVS4 simulations. 

In the present work, we mimic this approach, although we do not build our own distribution of semimajor axis jumps and encounter times from studies of Neptune encounters. Instead, we employ distribution functions designed to replicate the distributions of encounter times and semimajor axis jumps provided in \citet{nes16}. To broadly replicate the semimajor axis jump distribution in Figure 4 of \citet{nes16}, we draw from a Gaussian distribution where:

\begin{equation}
f(\Delta a) = \frac{1}{\sqrt{2\sigma^2\pi}}\exp{-\frac{(\Delta a - \mu)^2}{2\sigma^2}}
\end{equation}

To generate our jumps we assume that half of all jumps originate from a negatively centered Gaussian distribution. In these cases, we set $\mu=-2.5\times10^{-3}$ AU and $\sigma=1.8\times10^{-3}$ AU. For the other half of jumps we draw from a positively centered Gaussian. In these cases, we keep $\sigma$ at the same value and change the sign of $\mu$. Finally, the entire distribution is clipped at $\pm5\times10^{-3}$ AU to eliminate excessively large jumps not seen in the distributions provided in \citet{nes16}. This results in the distribution of semimajor axis jumps shown in Figure 1A, which generally matches the empirical form seen in \citet{nes16}. 

With a semimajor axis jump distribution created, we next need to set the total number of jumps. In each of our simulations, we set the total number of jumps equal to 20,000. This is the approximate amount of encounters expected for Neptune if it migrates through a disk containing 2000 Pluto-mass objects, which is at the lower end of the initial number of Pluto-mass objects predicted in \citet{nes16}. 

Finally, to determine the times at which semimajor axis jumps are given to Neptune, we construct an encounter time distribution function. First, we assume that until $t=\tau_{1} / 10$, the encounter times are uniformly distributed. After this point, we assume that they fall off as $t^{-1.15}$ until $t=3\tau_{2}$ as described in \citet{nes16}. This encounter time distribution is shown in Figure 1B for $\tau_{1}=30$ Myrs and $\tau_{2}=100$ Myrs. It can be directly compared to Figure 4 of \citet{nes16} and provides an approximate match to their encounter times. Regardless of the values of values of $\tau_{1}$ and $\tau_{2}$ we employ, the number of encounters is fixed at 20,000.

\begin{figure}
\centering
\includegraphics[scale=0.43]{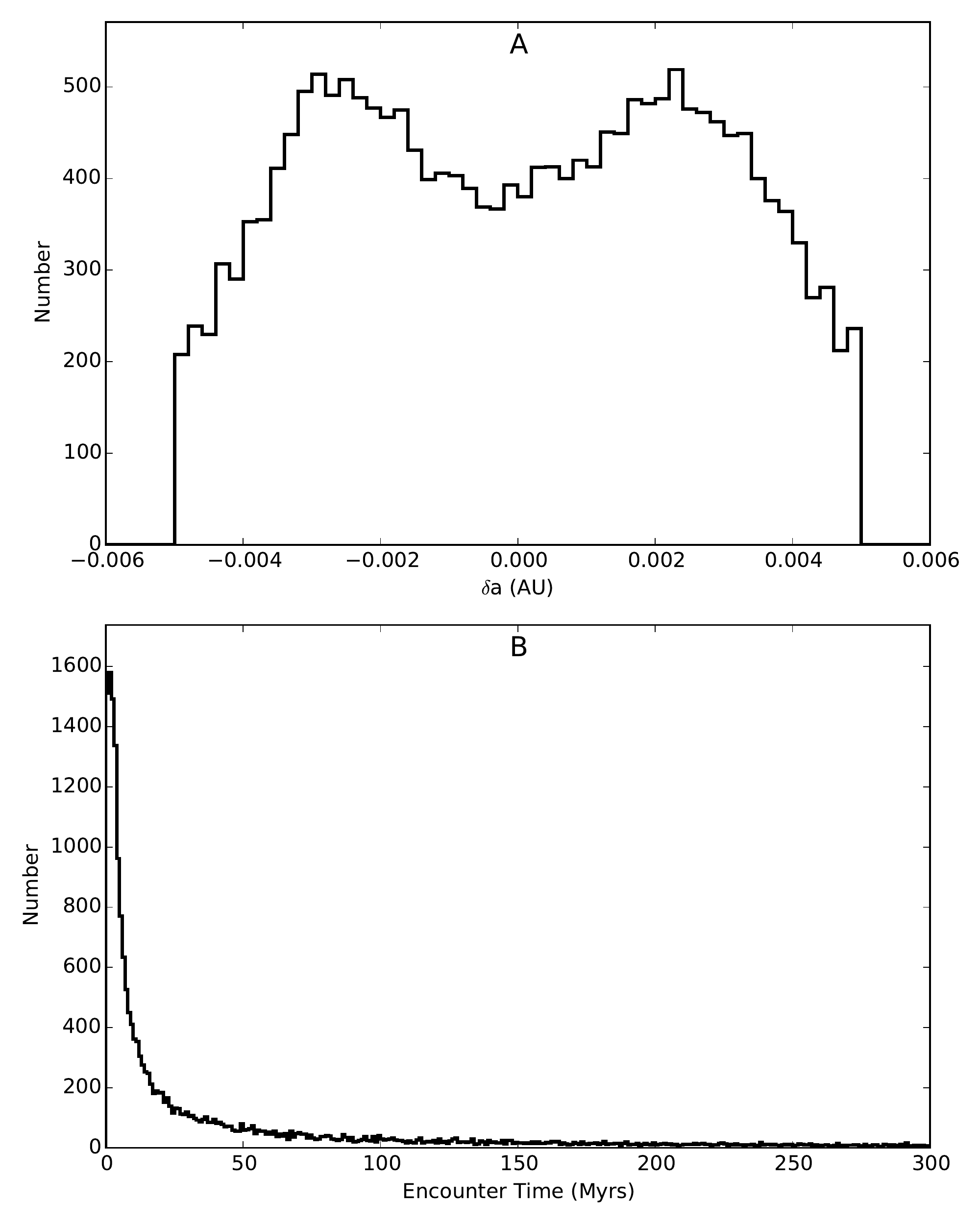}
\caption{{\bf A}: Distribution of Neptunian semimajor axis jumps used for our simulations with grainy migration. {\bf B}: Distribution of semimajor axis jump times used for our $\tau_{1} = 30$ Myrs, $\tau_{2} = 100$ Myrs simulation with grainy migration. }
\end{figure}

In total, we perform 4 different simulations. These simulations use two different migration timescales for Neptune: a slow and fast one. They also employ two different migration styles for Neptune: smooth and grainy.  In every simulation, Neptune experiences a 0.5 AU jump in semimajor axis at $\sim$28 AU, and it is given a post-jump eccentricity of 0.075, which slowly damps over time. Our simulations span the approximate range of migration parameters preferred by the recent works of \citet{nes15a, nes15b} and \citet{nes16}. Table 1 summarizes our simulations. Our simulation names describe the migration employed and stand for ``smooth slow'' (SmS), ``smooth fast'' (SmF), ``grainy slow'' (GS), ``grainy fast'' (GF). 

\begin{table}[htbp]
\centering
\begin{tabular}{c c c c c}
\hline

Run Name & $\tau_{1}$ & $\tau_{2}$ & Migration Style \\
 & (Myrs) & (Myrs) &  \\
\hline
SmS & 30 & 100 & Smooth\\
SmF & 10 & 30 & Smooth\\
GS & 30 & 100 & Grainy\\
GF & 10 & 30 & Grainy\\
\hline
\end{tabular}
\caption{The columns are: (1) the name of the simulation set, (2) the migration/damping timescale before Neptune's jump at 28 AU, (3) the migration/damping timescale after Neptune's jump, and (4) whether Neptune's migration is smooth or grainy.}
\end{table}

\section{Results}

The final eccentricities and semimajor axes of all surviving particles in our four simulations are shown in Figures 2A--D. Throughout this paper, we will plot the barycentric orbital elements of our particles. Because we are interested in the semimajor axis distribution of particles over very small ranges near resonance locations, we prefer barycentric elements. For orbits well beyond the planetary region, planetary perturbations on the Sun's position and velocity introduce a high-frequency jitter into heliocentric orbital elements that can hamper our analysis \citep{dones04}.

Several features are obvious in Figure 2. First, our runs with slow migration ($\tau_{1} = 30$ Myrs, $\tau_{2} = 100$ Myrs) result in a more anemic Kuiper belt than our runs with fast migration ($\tau_{1} = 10$ Myrs, $\tau_{2} = 30$ Myrs). Secondly, for particles with perihelia near 30 AU or beyond 40 AU, semimajor axes are clustered near the locations of mean motion resonances with Neptune. In the case of low perihelion orbits, their residence with Neptunian MMRs shields them from close encounters with Neptune, preserving their dynamical stability \citep{cohub65}. For orbits with perihelia beyond 40 AU, the clustering is due to perihelion-lifting that occurs when Kozai cycles operate within a MMR with Neptune \citep{gomes05}. Throughout this work, we will refer to orbits with $q>40$ AU as ``high-perihelion'' orbits. The range of simulation parameters explored in Figure 2 have been found to yield classical Kuiper belt orbital distributions that are generally consistent with observed detections \citep{nes15a, nes15b, nes16}, but high-perihelion distributions have not been studied as closely. These high-perihelion objects are the main focus of this work. (By eye, it appears that certain features of our simulated classical belts may differ somewhat from survey detections, such as significant resonant populations interior to the 3:2 MMR and a population of objects trailing the modern 2:1 MMR. However, this is beyond the scope of our current work, and we leave this to the subject of future studies.)

In Figure 3, we show the number of high-perihelion orbits as a function of their orbital period ratios with Neptune. Here the relative significance of the various mean motion resonances becomes clearer. Outside of resonances, Kozai cycles do not inflate the particles' perihelia, so the resonant populations are extremely prominent when we only consider high-perihelion particles. We see that regardless of the simulation, there is always a prominent population of high-perihelion particles near the 3:1 MMR. In addition, the 4:1 MMR also typically has a substantial population. In runs with grainy migration, we see that there are significant numbers of objects located near the 7:3 and 5:2 resonances as well, but their numbers are more suppressed relative to the 3:1 population if we only consider smooth migration scenarios. Finally, we note that the overall numbers of high-perihelion particles varies substantially from run to run. The fraction of surviving Kuiper belt objects with perihelia beyond 40 AU at the end of our simulations is 8.9\%, 3.7\%, 17.6\%, and 4.4\% for our SmS, SmF, GS, and GF simulations, respectively. Thus, the fraction of high-perihelion Kuiper belt objects seems to be enhanced by both grainy migration and slower migration timescales.

\begin{figure*}
\centering
\includegraphics[scale=0.44]{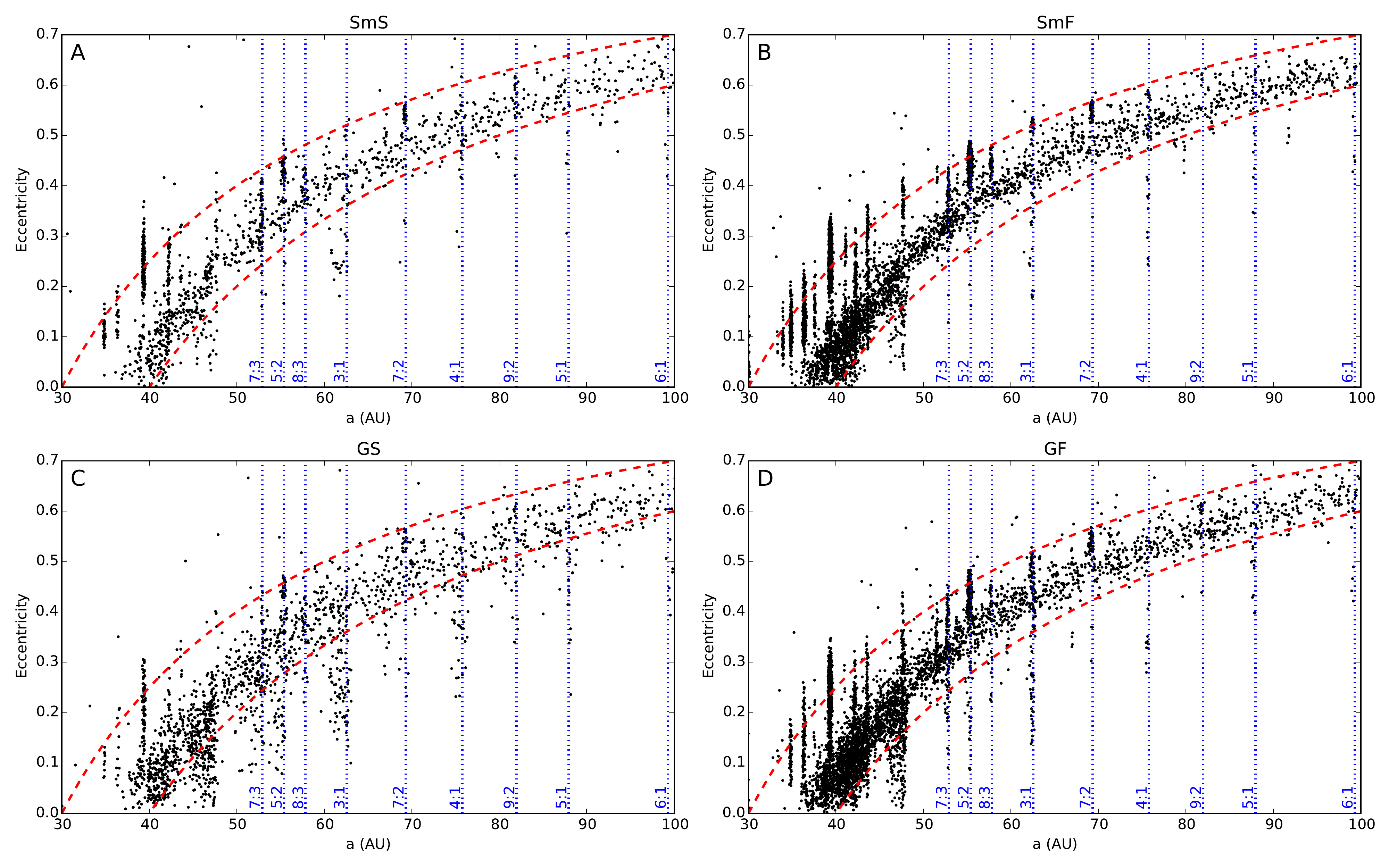}
\caption{Plots of our particles' final eccentricities vs semimajor axes for our SmS, SmF, GS, and GF simulations in panels A, B, C, and D, respectively. Red dashed lines mark $q=30$ AU and $q=40$ AU orbits. Blue dotted lines mark major mean motion resonances with Neptune.}
\end{figure*}

\begin{figure*}
\centering
\includegraphics[scale=0.44]{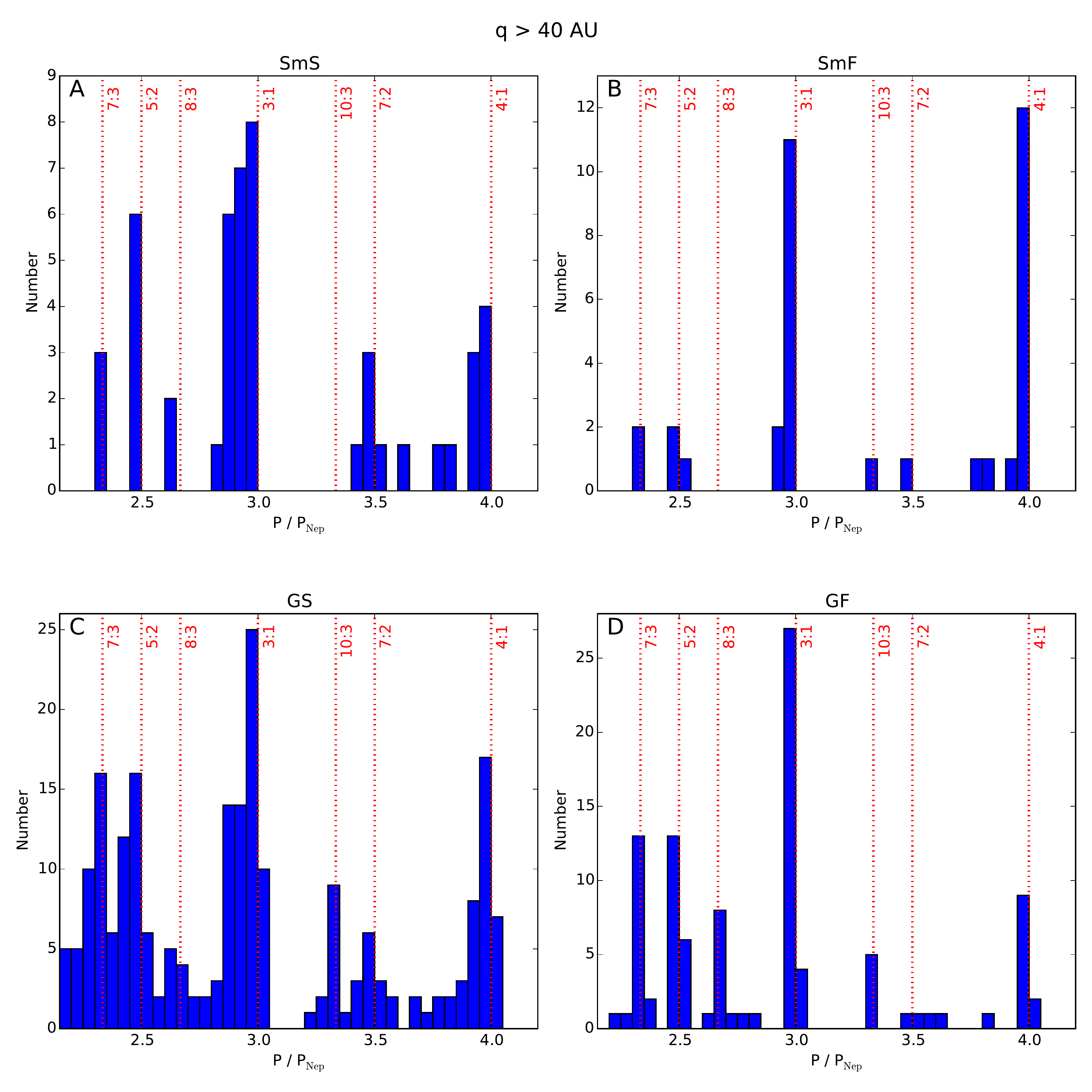}
\caption{The number of particles with $q>40$ AU is plotted as a function of their final orbital period ratio with Neptune. Significant mean motion resonances are marked with red dotted lines. The final states of our SmS, SmF, GS, and GF simulations are shown in panels A, B, C, and D, respectively.}
\end{figure*}

\subsection{Resonance Trails from Slow Migration}

In Figures 2 and 3, we also notice a qualitative difference between the simulations with slow migration and fast migration. For those with fast migration, the boundaries of the resonant populations are much more narrow. While resonances are still marked with population peaks in the the slow migration runs, the overall population distributions are smeared out. This is especially clear for the 3:1 and 4:1 resonances, since they are well-populated in each of our runs. This smearing occurs because when Kozai cycles inflate a particle's perihelion it becomes less dynamically coupled to Neptune, and it is easier to fall out of resonance \citep{gomes11}. If Neptune's semimajor axis is changing during this process, the object can leave the resonance permanently \citep{gomes05, gomes11}. Consequently, if Neptune is still migrating when Kozai cycles are operating, each resonance can leave a trail of particles as it moves outward.

This population of particles trailing the resonances only seems to be produced in certain Neptune migration scenarios. Before particles can fall out of resonance at high perihelion, they first must be captured into the resonance. Then they must have their inclinations increased to the point that Kozai cycles are activated \citep{gomes05}. Only after this will they attain high ($q>40$ AU) perihelia. Thus, the formation of a trail of particles just interior to a given resonance requires that the dynamical timescale for resonant capture and the initiation of Kozai cycles within the resonance be shorter than Neptune's migration timescale.

Figure 4 illustrates the difference in high-perihelion objects in our smooth and fast migration simulations. We first consider a particle within our GS simulation. In panel A, we plot the ratio of the particle's orbital period to Neptune's orbital period against the particle's perihelion. When the simulation first begins, Neptune is migrating with an $e$-folding time of 30 Myrs, and the particle is scattering off of the ice giants. Consequently, the particle's period changes rapidly. During this phase, Neptune's semimajor axis jumps by 0.5 AU at 28 AU and $t\simeq30$ Myrs, and Neptune then begins migrating with a slower timescale of 100 Myrs. Approximately 20 Myrs after Neptune's jump, the scattering particle is captured in the 5:2 resonance. This capture can be seen in panel B, as the critical resonant angle begins librating around the 70-Myr mark. Shortly after resonance capture occurs, the particle's perihelion is driven outward by the Kozai mechanism. Because Neptune is still migrating at this point and the particle's high perihelion weakens its coupling with Neptune, the particle falls out of resonance at high perihelion. This can be seen with the resumption of resonant angle circulation in panel C. With the particle out of resonance, Neptune continues to migrate, and the 5:2 resonance leaves the particle behind, frozen at high perihelion and an orbital period slightly less than 2.5 times Neptune's. 

Next we look at the evolution of a particle from our SmF simulation in panel C of Figure 4. Like the previous particle, this one begins by scattering off of the giant planets, and it is also eventually captured into the 5:2 resonance within the first 100 Myrs. However, in this particular simulation, Neptune is migrating outward faster. At about $t=130$ Myrs Neptune has completely finished migrating, and the particle is still in resonance, as indicated by the continued libration seen in panel D. From this point on, when Kozai cycles drive the particle to high perihelion, its semimajor axis will always remain near the 5:2 MMR since Neptune's semimajor axis is not evolving. Thus, with faster migration timescales, there is a much shorter opportunity for resonances to develop large fossilized populations of objects that fall out and trail the resonances. 

\begin{figure*}
\centering
\includegraphics[scale=0.44]{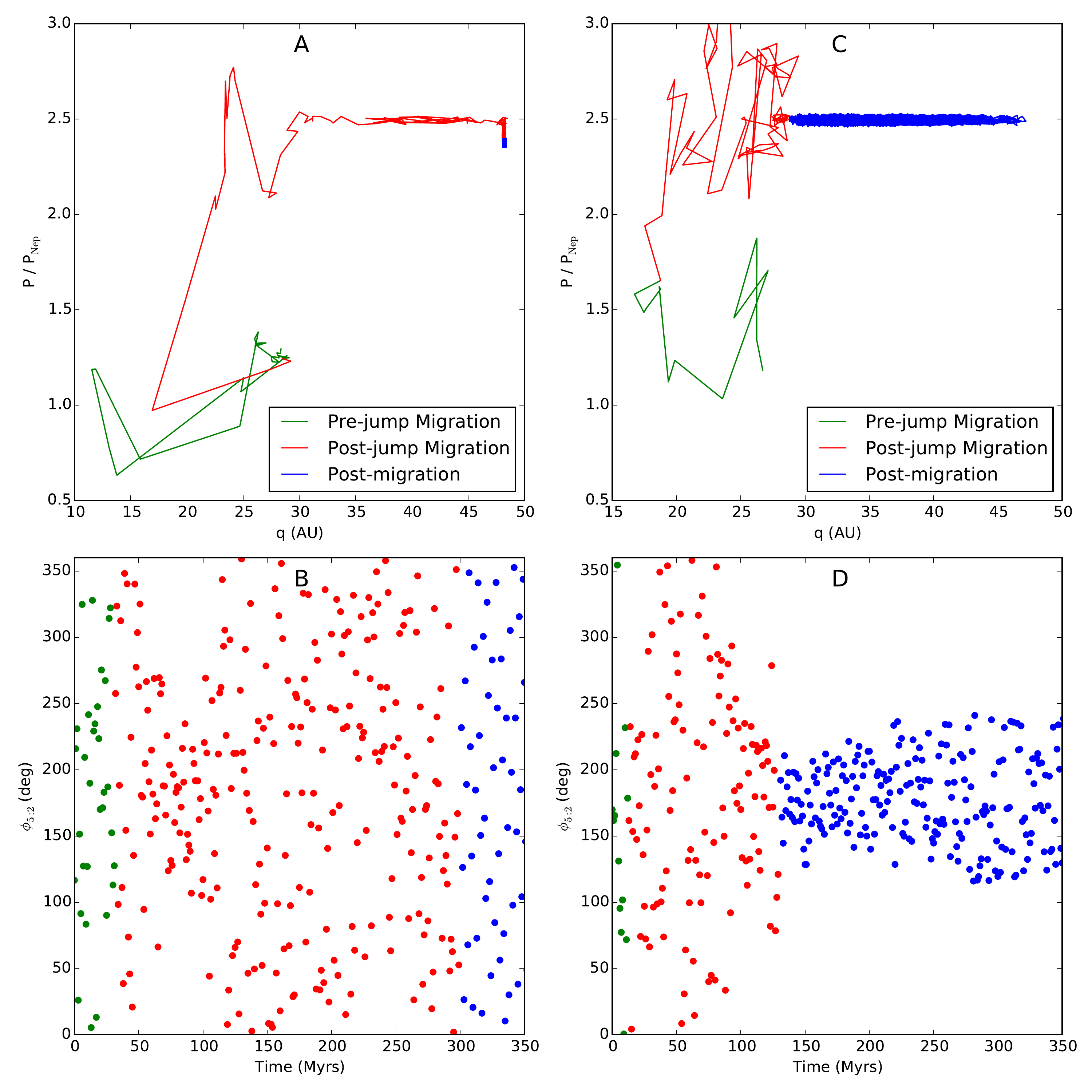}
\caption{{\bf A}: The 4-Gyr orbital evolution of a GS simulation particle whose perihelion is driven beyond 40 AU within the 5:2 resonance with Neptune. The ratio of the particle's orbital period to Neptune's is plotted against its perihelion. The green line marks the evolution that occurs before Neptune's jump at 28 AU. The red line marks the evolution that occurs after Neptune's jump but before Neptune stops migrating, and the blue line marks the evolution that occurs after Neptune stops migrating. {\bf B}: The critical resonant angle for the 5:2 resonance with Neptune as a function of time for the particle from Panel A. The color of the data points correspond to the colored times marked in Panel A. (Only the first 350 Myrs of evolution are shown in this panel to highlight the libration near $t=$ 50--100 Myrs.) {\bf C}: The 4-Gyr orbital evolution of a GF simulation particle whose perihelion is driven beyond 40 AU within the 5:2 resonance with Neptune. The ratio of the particle's orbital period to Neptune's is plotted against its perihelion. The line coloring is analogous to panel A. {\bf D}: The critical resonant angle for the 5:2 resonance with Neptune as a function of time for the particle from Panel C. The color of the data points correspond to the colored times marked in Panel C.}
\end{figure*}

\subsubsection{Trailing Population of the 3:1}

Of all the resonances that show trailing populations, the 3:1 MMR may be the most useful for several reasons. First, it is well-separated from other major mean motion resonances (unlike the 5:2, which is closely flanked by the 7:3 and 8:3 resonances), so there is less chance to confuse the influences of different resonances. Even in the GS simulation, which has the largest resonance trails, it is easy to separate the 3:1 objects from the rest of the high-perihelion Kuiper belt. In addition, it is well-populated in each of our simulations, so a statistically useful distribution of orbits can be attained. Finally, it is closer to the Sun than our other consistently well-populated, well-separated resonance, the 4:1, so objects in the 3:1 are about twice as easy to detect according to \citet{shep16}.

In Figure 5, we plot the last time that particles near the 3:1 had perihelia below 40 AU against their final semimajor axes. It should be noted that the semimajor axes shown in the figure are calculated from the simulated particles' orbital period ratios with Neptune. In these calculations, we always assume that Neptune's final semimajor axis is 30.11 AU, so we can directly compare with the actual solar system. In reality, our simulations all finish with slightly different final orbits for Neptune, which would slightly alter the location of the 3:1 MMR and make it more confusing to compare our simulations against each other.

By plotting the last time that particles have $q<40$ AU in Figure 5, we are effectively marking the time at which they are last strongly coupled to Neptune's dynamics. After this time, the particles have perihelia above 40 AU for the rest of the simulation and are less coupled to Neptune. One can see in Figure 5, that there is a strong correlation between this time and a particle's final semimajor axis. For particles very near the 3:1 MMR (which is at 62.6 AU), many of them have had their perihelia below 40 AU during the last 3 Gyrs. This is because these particles are still occasionally undergoing Kozai cycles within the 3:1 MMR. (These particles are not guaranteed to have had extremely recent excursions to $q<40$ AU, since resonance libration and Kozai oscillations can temporarily stop for orbits near MMRs \citep{gomes11}. Nonetheless, they are ``resonantly active'' over timescales comparable to the solar system's age.) Meanwhile, for particles just below $\sim$62 AU, there are no instances of recent excursions to $q<40$ AU. These are particles that were trapped in the 3:1 MMR while Neptune was still migrating, and during a Kozai cycle to high-perihelion, they were dropped by the 3:1 MMR as Neptune continued to migrate outward. At that point, their orbital evolution becomes essentially frozen, and they become a fossilized relic of a past location of the 3:1 MMR. As a result, their final semimajor axes are simply a function of when they dropped out of resonance during Neptune's migration.

One only sees this trend strongly in the simulations featuring a ``slow'' migration speed for Neptune: SmS and GS. In these simulations, $\sim$50\% of all particles near the 3:1 have semimajor axes below 62 AU, and all of these particles fell out of resonance while Neptune was still migrating. In the ``fast'' simulations (SmF and GF), there is only 1 particle with a semimajor axis well below 62 AU. The reason for this is that Neptune is migrating faster than particles can be captured into the 3:1 MMR and initiate Kozai cycling, which is required to fall out of resonance and become fossilized at high perihelion. There are instances where particles are dropped from the 3:1 MMR before $t=100$ Myrs, but even then, Neptune has already migrated most of the way to its current orbit, so the difference in final semimajor axes of these dropped particles and the resonantly active particles is much less dramatic than in our slow migration simulations. 

We also show the barycentric semimajor axes of all detected TNOs near the 3:1 MMR with $q>40$ AU. Until very recently, only one such object was known, but \citet{shep16} announced the detections of 3 more, and the Pan-STARRS survey added another detection \citep{weryk16}. While 4 of the 5 known 3:1 high-perihelion objects are very near the resonance center and consistent with resonantly active orbits, 2014 JM$_{80}$ sits near the inner edge of our resonantly active particles. The nature of 2015 KH$_{162}$ is less ambiguous. This object has a semimajor axis of 61.7 AU compared to the nominal 3:1 MMR location of 62.6 AU. Thus, it sits 0.9 AU closer to the Sun than the 3:1 MMR, and its orbital period is over 2\% shorter than objects trapped in the 3:1 MMR. For 2015 KH$_{162}$ to be in resonance, Neptune's semimajor axis would need to be $\sim$29.7 AU. In our simulations, all of the test particles with orbits near a semimajor axis of 61.7 AU were decoupled from the planets within the first $\sim$200 Myrs. 2015 KH$_{162}$ is likely an example of an object released from the 3:1 MMR before Neptune completed its migration. The only other clear example of a known object like this is 2004 XR$_{190}$, which is hypothesized to have fallen out of the 8:3 MMR during Neptune's migration \citep{gomes11}. In the case of 2004 XR$_{190}$, it sits about 0.3 AU closer to the Sun than the 8:3 MMR center, and its orbital period is about 1\% shorter than objects in the 8:3. Since 2015 KH$_{162}$ is even further from its MMR than 2004 XR$_{190}$, 2015 KH$_{162}$ likely fell out of resonance even earlier in Neptune's migration than 2004 XR$_{190}$.

\begin{figure*}
\centering
\includegraphics[scale=0.44]{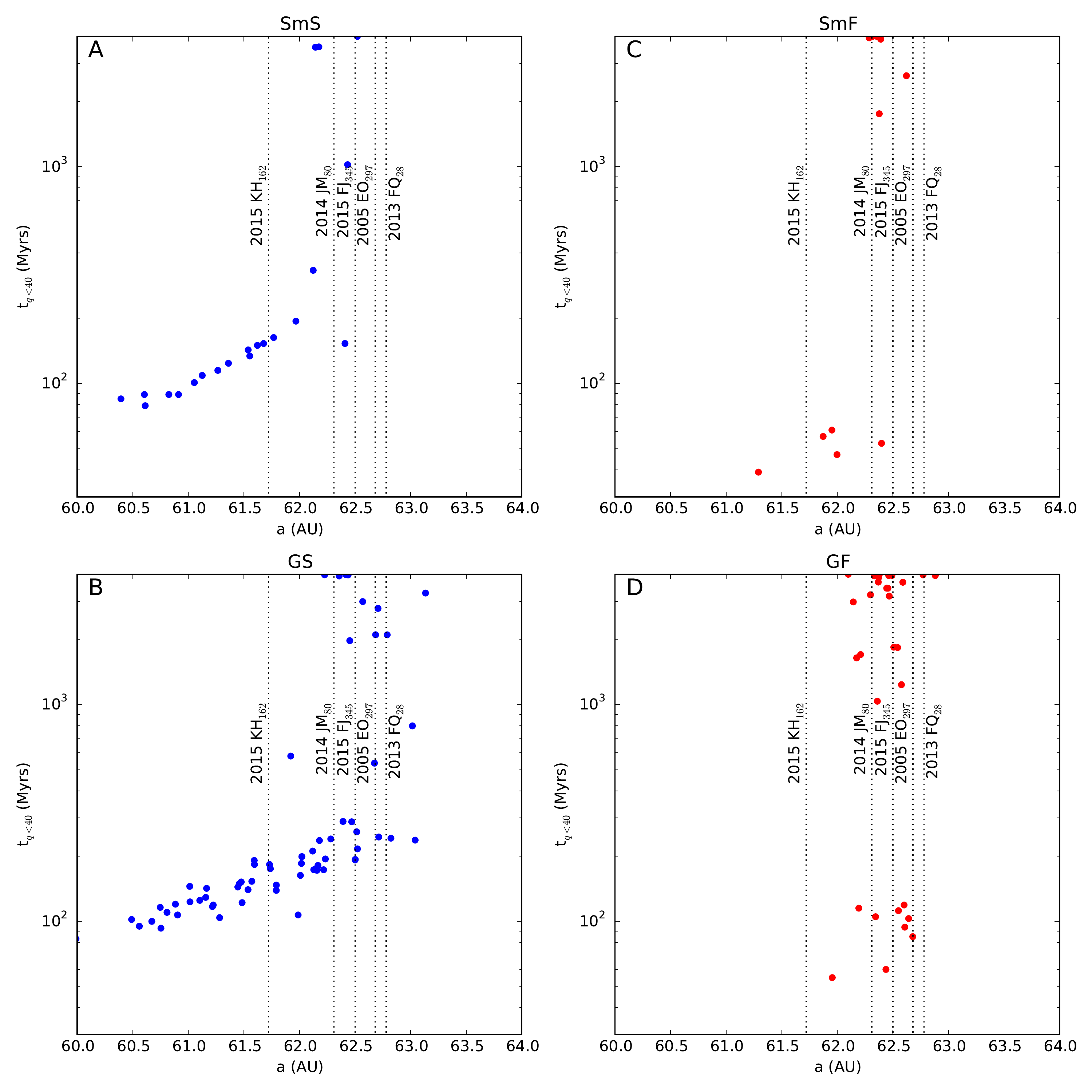}
\caption{For particles with perihelia beyond 40 AU and located near the 3:1 MMR with Neptune at $a=62.6$ AU, the last times at which they had perihelia inside 40 AU are plotted against their final semimajor axes. The semimajor axes of the five known TNOs near the 3:1 MMR with $q>40$ AU are marked with vertical dotted lines. Simulation results for our SmS, SmF, GS, and GF runs are shown in panels A, B, C, and D, respectively.}
\end{figure*}

\subsubsection{Trailing Populations of Other Resonances}

The 3:1 resonance is not the only one that displays a trailing population. In Figure 6, we also look at the 5:2 and 4:1 resonances, which often have prominent high-perihelion populations and a final semimajor axis distribution that varies with different migration scenarios. Figure 6 shows the semimajor axis distributions for high-perihelion objects within $\pm$2 AU of the 5:2, 3:1, and 4:1 resonances. We notice that with grainy migration, the 5:2 and 4:1 resonances display trends that are qualitatively similar to the 3:1. Namely, they have extended trails for slow migration and do not otherwise. When Neptune's migration is smooth, however, these trails are not seen. Objects in these resonances seem to be able to stay coupled to Neptune unless sudden small jumps occur in Neptune's semimajor axis evolution. (It should be noted that the handful of objects that are over 1 AU exterior to the 5:2 resonance actually fell out of the 8:3 MMR early in Neptune's migration.)

Thus, while the 5:2 and 4:1 resonances may provide diagnostics on the graininess of Neptune's migration, the 3:1 resonance may hold the best constraints on Neptune's migration timescale. We see that when Neptune migrates slowly ($\tau_{1} = 30$ Myrs, $\tau_{2} = 100$ Myrs), at least $\sim$40--50\% of high-perihelion objects near the 3:1 MMR are at least 1 AU interior to the resonance center. Meanwhile when Neptune migrates fast, this number is under 10\%. 

\begin{figure*}
\centering
\includegraphics[scale=0.44]{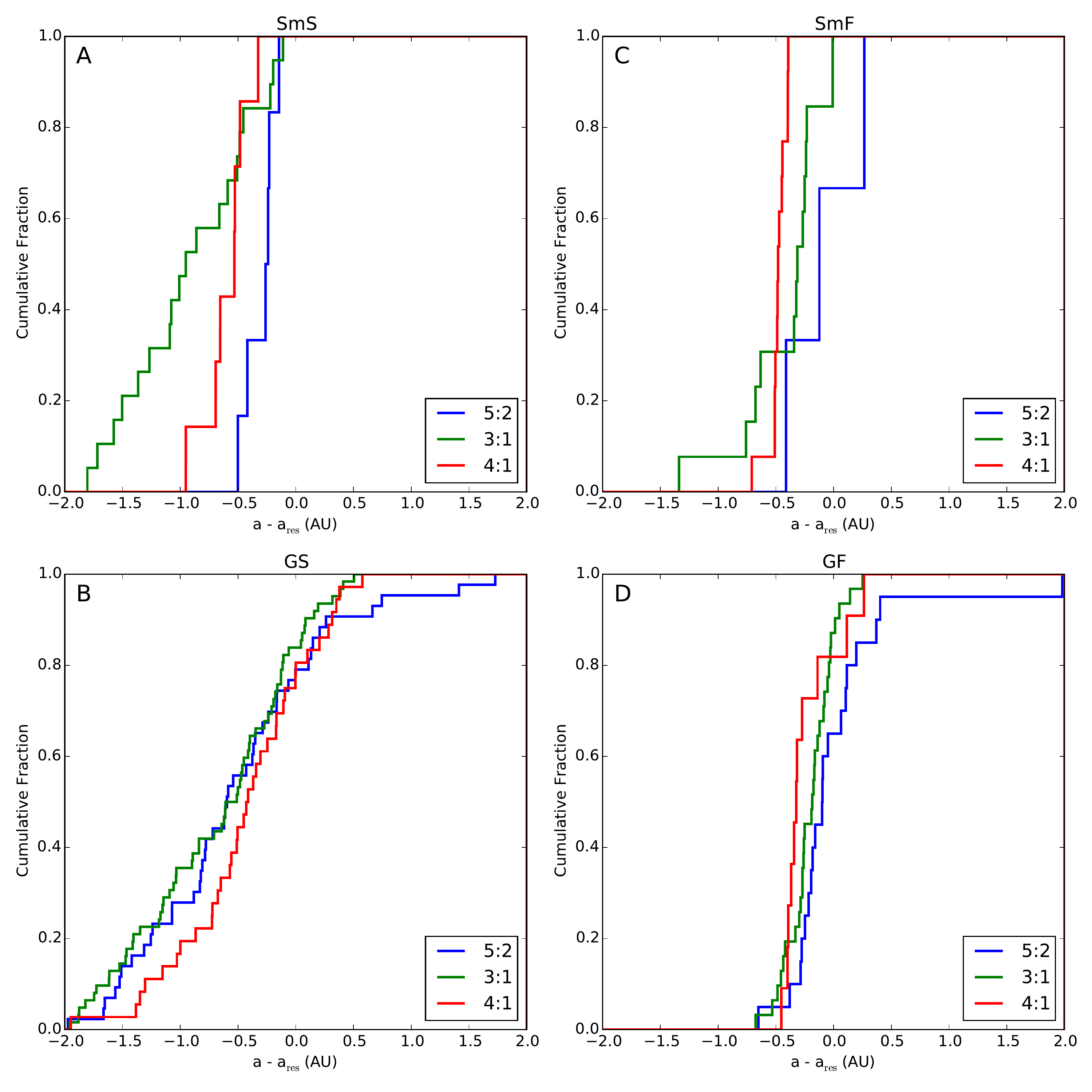}
\caption{For particles with perihelia beyond 40 AU, the cumulative distributions of particle semimajor axes are shown within $\pm$2 AU of the 5:2 ({\it blue}), 3:1 ({\it green}), and 4:1 ({\it red}) resonances with Neptune. Results from our SmS, SmF, GS, and GF simulations are shown in panels A, B, C, and D, respectively.}
\end{figure*}

\subsection{Relative Resonance Populations}

Our simulations also provide a prediction on the relative numbers of objects near each resonance. Even though our slow migration simulations have a smeared out semimajor axis distribution, we can still associate particles fairly accurately with their dynamically relevant resonance by simply searching for the nearest major resonance to it in terms of the particle's orbital period. This is done in Figure 7, and we use it to compare the relative populations of each major resonance. One major feature that is clear in Figure 7 is that grainy migration enhances the number of high-perihelion objects near the 7:3, 5:2, and 8:3 resonances. With grainy migration these populations (7:3, 5:2, and 8:3) sum to $\sim$150\% of the 3:1 population, whereas with smooth migration, the combined population of these three resonances is 40--50\% of the 3:1 population. The relative size of the 4:1 population also varies throughout our simulations between 40\% and 115\% of the 3:1 population, but there is not an obvious trend between simulations.

\begin{figure*}
\centering
\includegraphics[scale=0.44]{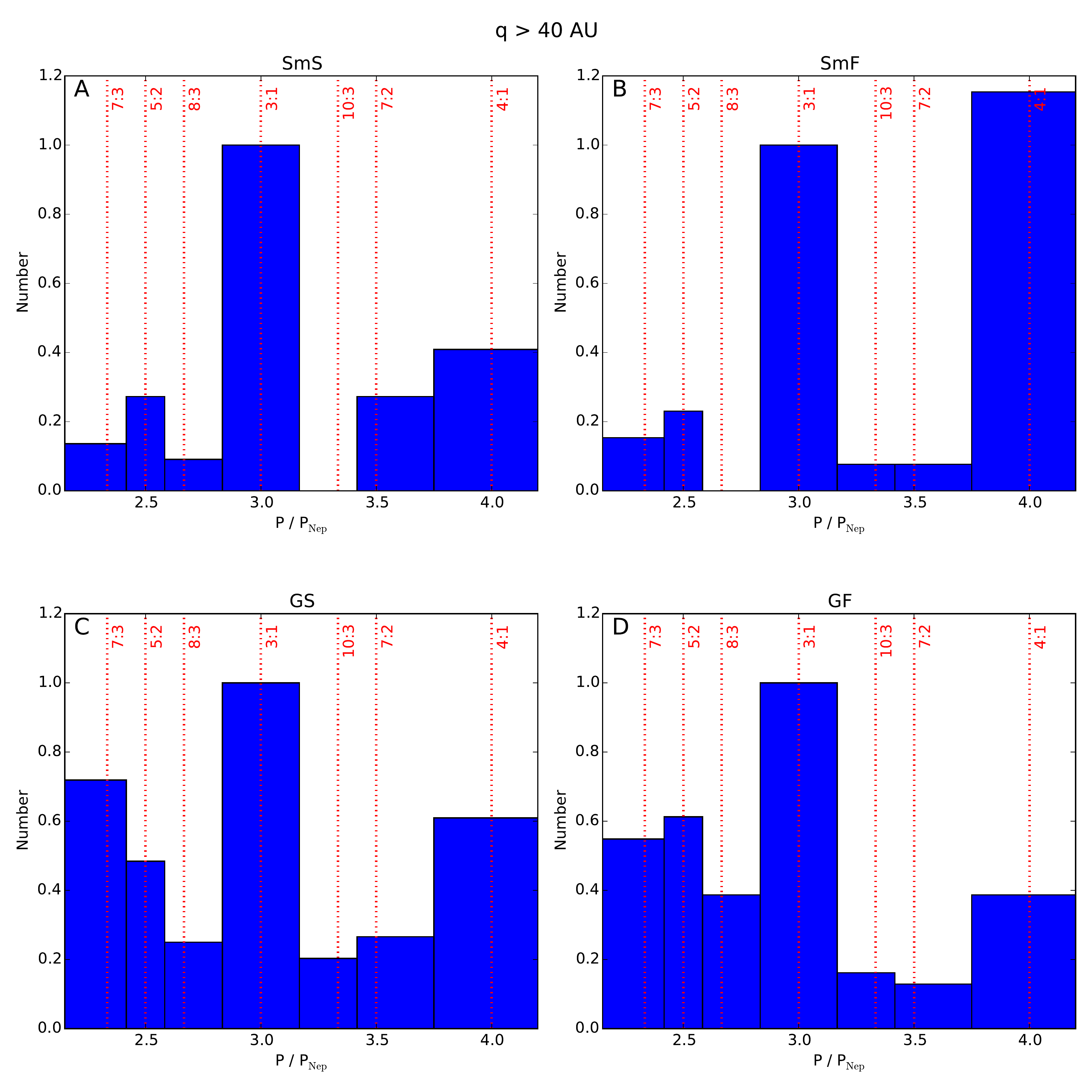}
\caption{The final numbers of particles with $q>40$ AU is shown between the 7:3 and 4:1 resonances for each of our simulations. Bins are divided halfway between each of the MMRs marked with red dotted lines. Results from our SmS, SmF, GS, and GF simulations are shown in panels A, B, C, and D, respectively.}
\end{figure*}

We can also look at the times that high-perihelion orbits are populated. We first do this in Figure 8A, where we display the distribution of times at which high-perihelion objects last had perihelia below 40 AU for each of our simulations. We can see that these distributions are fairly bimodal. There is one subpopulation of objects that last had $q<40$ AU when Neptune was still migrating. These trail the actual modern resonance locations and are fossilized at high perihelia after being decoupled from a migrating Neptune during Kozai cycles. A second subpopulation has often had excursions to perihelia below 40 AU in the last Gyr. This second group represents objects that were captured after Neptune stopped migrating or never fell out of resonance before Neptune stopped migrating. Because this population still resides near the centers of the resonances, they can still regularly undergo Kozai cycles and move back and forth across the $q=40$ AU boundary. Thus, these are still dynamically active objects that undergo Kozai oscillations within MMRs.

\begin{figure}
\centering
\includegraphics[scale=0.43]{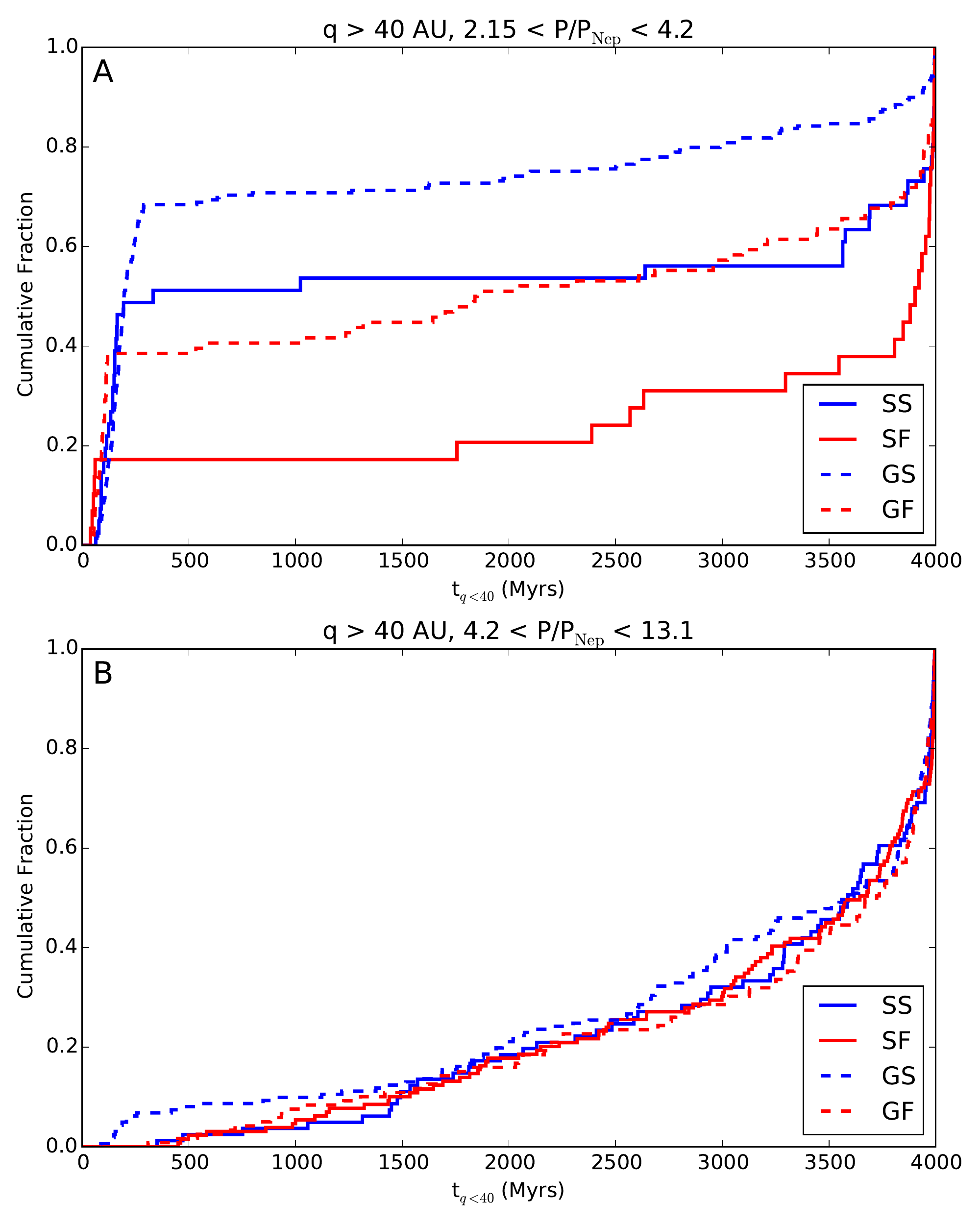}
\caption{Cumulative distributions are shown of the last time that each particle had a perihelion value inside 40 AU. Distributions are shown for our SmS ({\it blue solid}), SmF ({\it red solid}), GS ({\it blue dashed}), and GF ({\it red dashed}) simulations. {\bf A}: These distributions are for particles with a final perihelion above 40 AU and a final orbital period that is between 2.15 and 4.2 times that of Neptune. {\bf B}: These distributions are for particles with a final perihelion above 40 AU and a final orbital period that is between 4.2 and 13.1 times that of Neptune. }
\end{figure}

\subsection{Resonant Populations Beyond the 4:1 (at 76 AU)}

Up until now we have only considered the population near or closer than the 4:1 resonance. We of course also generate high-perihelion populations for resonances more distant than the 4:1, which we now consider here. We begin in Figure 8B by examining the times that these particles last had perihelia below 40 AU. Here we find a very different distribution of times compared to the closer resonant populations. Almost all high-perihelion objects beyond the 4:1 have spent time at low perihelion values since Neptune stopped migrating. This tells us that these particles generally do not fall out of resonance while Neptune is still migrating. More accurately, this indicates that the timescale for resonance capture and Kozai perihelion-lifting is typically longer than the timescale for Neptune's migration even in our slow simulations. If we had chosen still longer migration timescales of hundreds of Myrs, we would likely find trailing populations for some of these resonances as well. 

To estimate the length of time that it takes for a particle to become trapped in a given resonance and then have its perihelion lifted beyond 40 AU, we search all of our simulations for any particle that ever attains $q>40$ AU (even if it eventually reverts to $q<40$ AU and/or does not survive until the simulation's end). For each of these particles we record the time at which it first attains $q>40$ AU, and we also record its orbital period ratio with Neptune when it first makes the transition to $q>40$ AU. Based on the particle's orbital period ratio, we assign it to the 7:3, 5:2, 8:3, 3:1, 4:1, 5:1, or 6:1 MMR (we do not consider period ratios below 2.2 or above 6.2). In this way, we compile a distribution of times for perihelion-lifting for each of the resonances mentioned above. 

These distributions are shown in Figure 9. Here we see that perihelion-lifting is faster in the 3:1 MMR than in any of the other resonances. Particles in this resonance attain $q>40$ AU in a median time of 301 Myrs. Thus, half of the objects in the 3:1 that reach $q>40$ AU do so before Neptune has stopped migrating in our slow migration simulations. This is one reason why the 3:1 MMR population provides the strongest signature of Neptune's migration. In spite of the short median perihelion-lifting time in the 3:1, $\sim$90\% of these objects do not attain $q>40$ AU until after 100 Myrs, which is nearly the time at which Neptune stops migrating in our fast migration runs. This is why high-perihelion objects near the 3:1 do not exhibit a tail to smaller semimajor axes in our fast migration runs and why the distribution of objects near the 3:1 provides the tightest constraint on Neptune's migration timescale. Perihelion-lifting in the 7:3 is nearly as fast as the 3:1, but as seen in Figure 7, grainy migration is necessary for a significant number of these objects to fall out of resonance. The existence or absence of a resonance trail near the 7:3 can constrain the graininess of Neptune's migration.

Figure 9 also illustrates another important point. Since perihelion-lifting simply does not occur during the first 100 Myrs of our simulations, the ``jump'' of Neptune at 28 AU (as well as its pre-jump migration) is not constrained by the high-perihelion population of the Kuiper belt. In our GF and SmF simulations, this jump takes place at $t\simeq10$ Myrs, and in our GS and SmS simulations, the jump occurs at $t\simeq30$ Myrs. Both of these times are well before any objects have attained high perihelia. Thus, although we include a jump in Neptune's migration because it is well-motivated in recent works \citep{nesmorb12, nes16}, our high-perihelion results are not sensitive to it.

Beyond the 3:1 resonance, Figure 9 also shows that the perihelion-lifting timescale steadily increases with distance from the Sun for the N:1 resonances. In the 6:1 MMR, the median time at which particles attain $q>40$ AU is 1 Gyr, and less than $\sim$10\% of particles attain $q>40$ AU before Neptune stops migrating even in our slow migration simulations. This explains why, as seen in Figure 8, the vast majority of the high-perihelion particles beyond the 4:1 MMR have had $q<40$ AU at some point in the last 3 Gyrs; they are all still resonantly interacting with Neptune today. Thus, these more distant resonances should have few, if any, fossilized high-perihelion objects near them.

\begin{figure}
\centering
\includegraphics[scale=0.43]{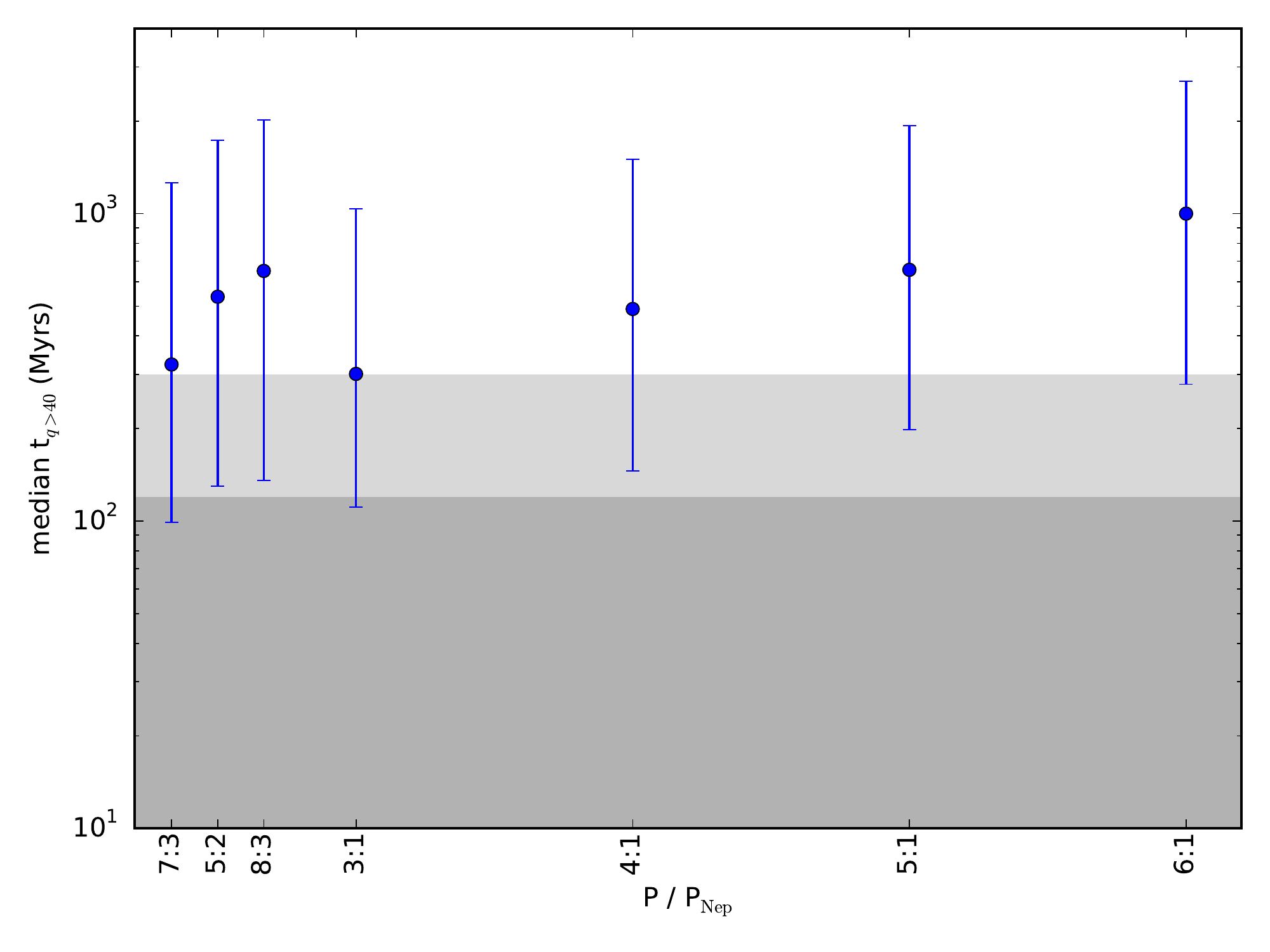}
\caption{The median time at which any particle first attains $q>40$ AU is plotted as a function of the nearest resonance when this occurs. The error bars mark the times at which 10\% and 90\% of all particles have attained $q>40$ AU within a given resonance. The dark gray shaded region mark times during which Neptune is still migrating in our fast migration simulations (SmF and GF). The dark {\it and} light gray shaded regions mark times during which Neptune is still migrating in our slow migration simulations (SmS and GS). }
\end{figure}

In general we find that the total numbers of high-perihelion particles steadily fall off beyond the 4:1 resonance. As a result, the numbers of objects populating these more distant resonances are less statistically significant. However, because these particles are not dynamically fossilized during Neptune's migration and have interacted more recently with Neptune, they represent a more modern population that is less tied to Neptune's early migration. Because of this, we elect to co-add our four simulation's particles to better study the relative prominence of each resonant population. This is done in Figure 10, where we show the number of particles as a function of their orbital period ratio with Neptune. One striking feature in this figure is that the N:1 resonant populations dominate. For high-perihelion particles with periods between 5 and 13 times that of Neptune, we find that $\sim$50\% are found with an orbital period that is within 10\% of an N:1 resonance. We also see instances of Kozai perihelion-lifting occurring at N:2 and N:3 resonances (where N is very large), but these are not as efficient. This population fills in the gaps between N:1 populations. It is also obvious in  Figure 10 that the N:1 population falls off with increasing N. If we attempt to fit a power-law to these populations, we find that the population size of the N:1 resonances decreases with N$^{-1.4}$.  Meanwhile, the falloff of the rest of the high-perihelion population in this period range is best fit by a power-law with a -1.2 exponent.

\begin{figure}
\centering
\includegraphics[scale=0.43]{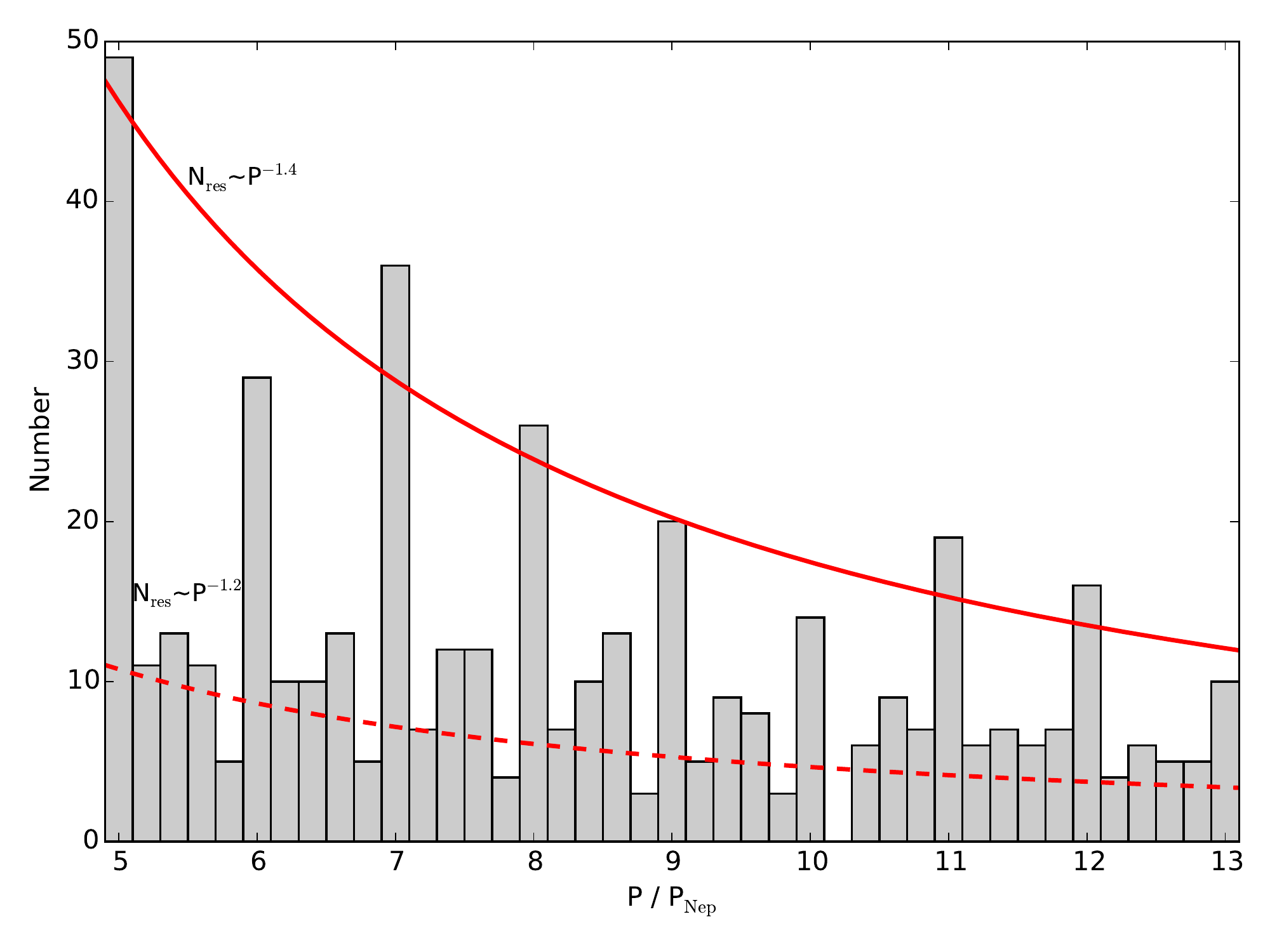}
\caption{The number of particles with $q>40$ AU is plotted as a function of their orbital period ratio with Neptune. These numbers are generated by co-adding our SmS, SmF, GS, and GF simulations. The solid red line marks a -1.4 power law, which is the best fit to the numbers of bodies within $\pm$0.1 of a N:1 period ratio with Neptune for for N between 5 and 13. The dashed red line marks a -1.2 power law, which is the best fit to the numbers of bodies further than $\pm$0.1 from a N:1 period ratio with Neptune for for N between 5 and 13.}
\end{figure}

\subsection{Inclination Distribution}

As mentioned previously, during Kozai cycles, an orbit's inclination and eccentricity oscillate exactly out of phase with each other so that the quantity $\sqrt{1-e^2}\cos{i}$ is conserved. Therefore, when the perihelia of TNOs are raised by Kozai cycles (and eccentricity is decreased) the orbital inclinations of these objects must go up as well. Based on this we expect high-perihelion objects to have a substantially hotter inclination distribution than the rest of the Kuiper belt. In Figure 11, we compare the inclinations of objects with $q>40$ AU to those with $q<40$ AU for all objects with orbital period between 2.1 and 13.1 times that of Neptune. To construct these distributions, we co-add all four of our simulations and weight each simulation by the inverse of the total number of simulation orbits, so that each simulation contributes equally. (In actuality, the distributions from each simulation are nearly identical.) As can be seen in this figure, the two distributions are radically different. For high-perihelion objects, the median inclination is $\sim$34$^{\circ}$, and 99\% of these objects have inclinations greater than 20$^{\circ}$. Moreover, inclinations can reach as high as $\sim$50$^{\circ}$. On the other hand, objects with perihelia below 40 AU are typically found at substantially lower inclinations. For these objects, the median inclination is 20.5$^{\circ}$, and 92\% of these objects have inclinations below 34$^{\circ}$, the median of the high-perihelion inclination distribution. (It should be noted that the classical Kozai mechanism for a circular orbit is not activated until the inclination exceeds $\sim$40$^{\circ}$. However, this is not true for these high-eccentricity orbits occupying MMRs. Hence, most of our orbital inclinations are below 40$^{\circ}$ even though they undergo Kozai cycles.) 

\begin{figure}
\centering
\includegraphics[scale=0.43]{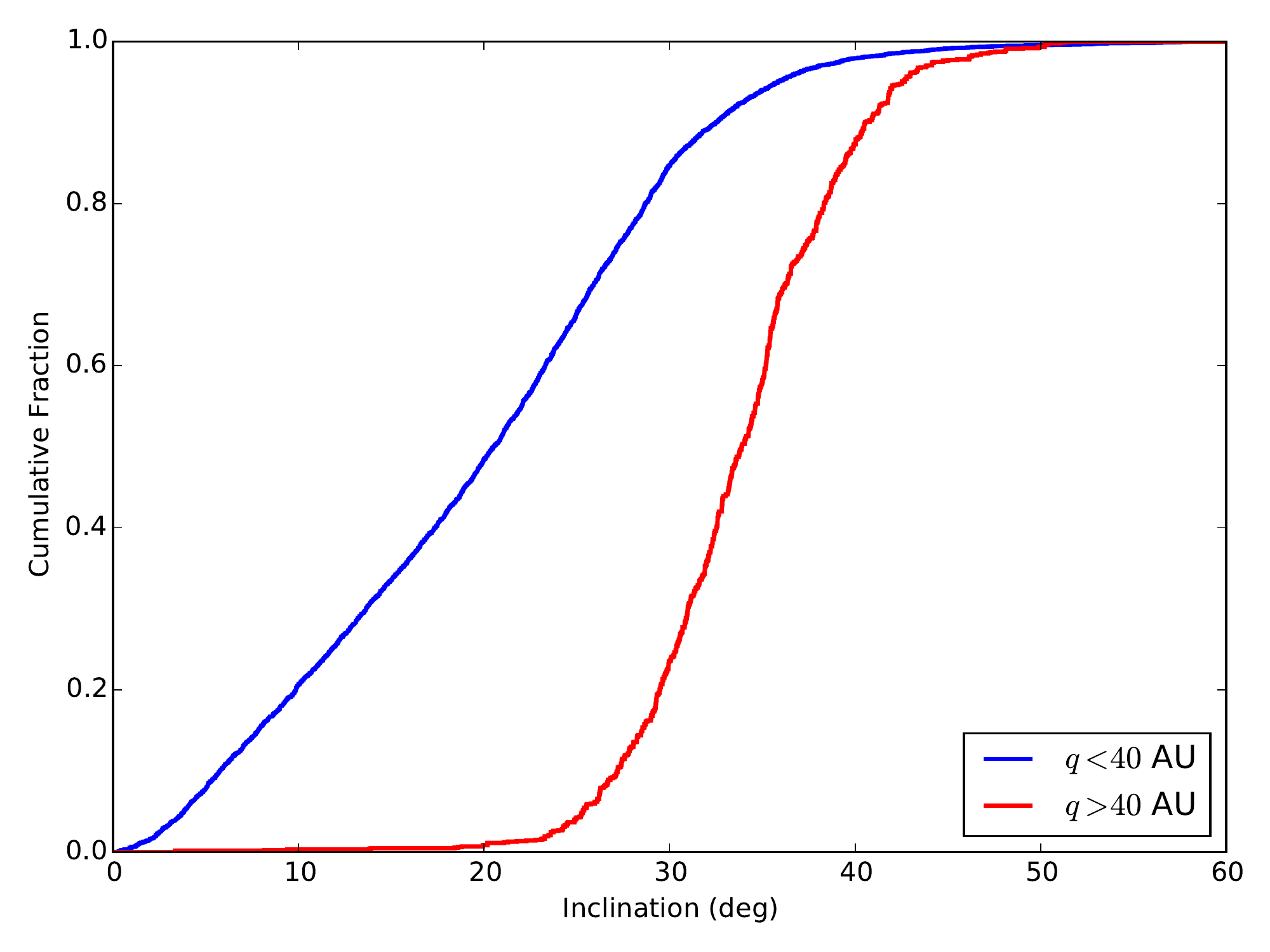}
\caption{The cumulative distribution of orbital inclinations for particles with orbital periods between 2.1 and 13.1 times that of Neptune. Two distributions are shown: one for particles with $q>40$ AU ({\it red}) and one for particles with $q<40$ AU ({\it blue}).}
\end{figure}

\subsection{Non-Resonant High-Perihelion Objects}

Although capture into a MMR with Neptune greatly enhances an object's probability of attaining a perihelion beyond 40 AU, it is not strictly necessary. A re-examination of Figure 2 shows numerous instances of objects with $q>40$ AU that do not seem to be associated with any major MMR. Typically, these have perihelia just beyond 40 AU, which serves to illustrate that 40 AU is not a fundamental perihelion value that objects are strictly prohibited from crossing without the aid of resonant or Kozai effects. We have investigated the evolution of many of these apparently non-resonant high-perihelion objects, and a typical example is shown in Figure 12. Here a particle is scattered to a semimajor axis of $\sim$70--75 AU within 100 Myrs, and it remains near this semimajor axis range for most of the rest of the simulation. Meanwhile, the perihelion of this object is near 35 AU for the first 1.3 Gyrs and eventually makes its way to 40--41 AU by the end of the simulation. When the particle makes notable shifts in perihelion, the orbital period ratio with Neptune is near $\sim$3.8, far from any of the major resonances explored in this work. The dynamics of the perihelion evolution are undoubtedly complex, but the evolution exhibits a behavior that looks qualitatively more akin to chaotic diffusion over Gyr-timescales rather than perihelion-lifting driven by Kozai cycles. Interestingly, despite not exhibiting Kozai-like behavior, we anecdotally find that these objects all have inclinations of at least 20--30$^{\circ}$ by the time they attain $q>40$ AU, similar to objects that attain $q>40$ AU via classic Kozai cycles within MMRs. 

\begin{figure}
\centering
\includegraphics[scale=0.43]{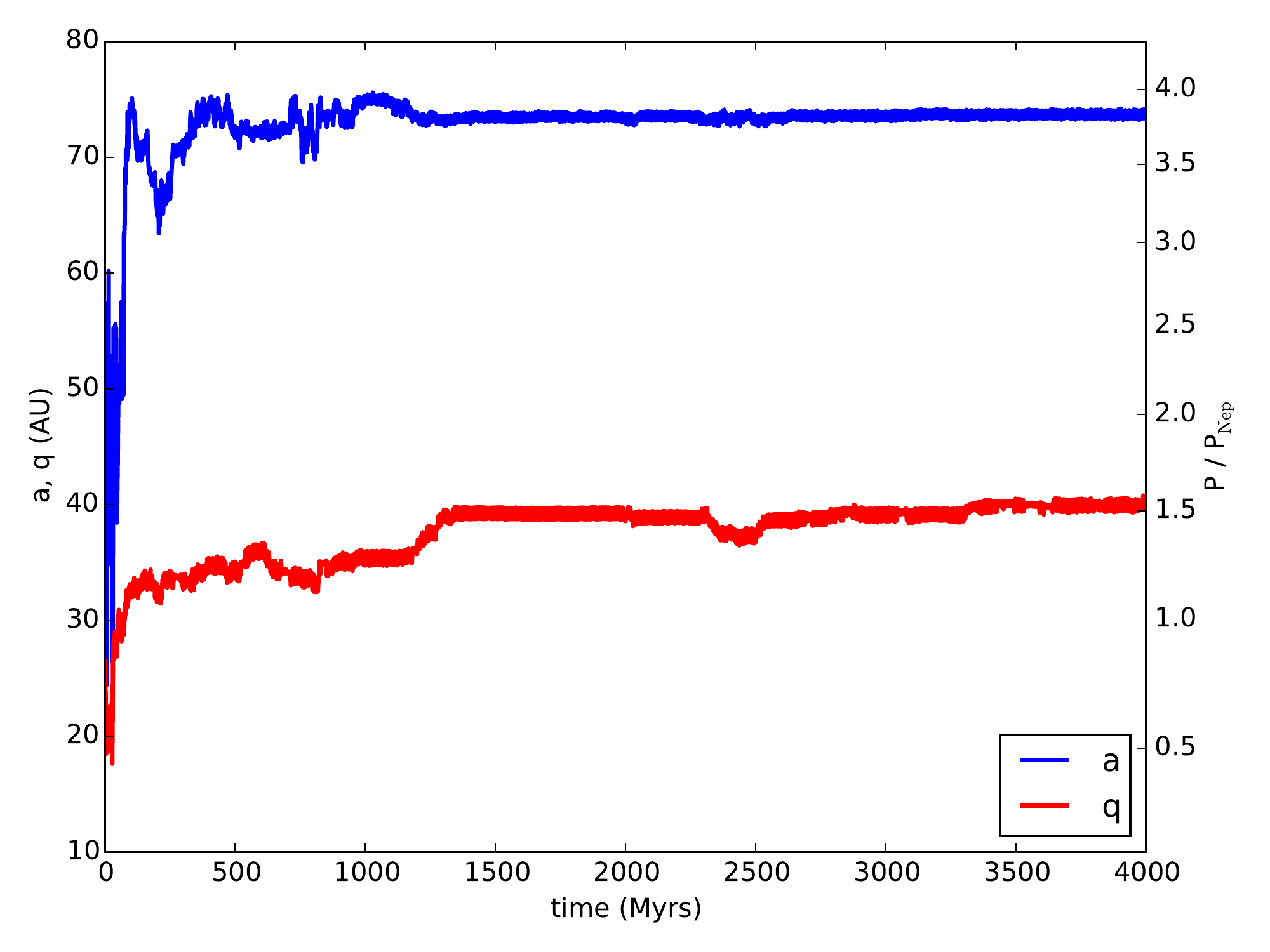}
\caption{The semimajor axis ({\it blue}) and perihelion ({\it red}) of a particle from our SmS simulation is plotted against time. The orbital period ratio of the particle with Neptune is given by the right $y$-axis.}
\end{figure}

Although evolutionary paths similar to the one shown in Figure 12 are rare, they are seen in every simulation. This type of behavior may explain the origins of known high-perihelion, moderate eccentricity objects that are not obviously linked to a MMR, such as 2014 QR$_{441}$ ($q=42.6$ AU, $P / P_{\rm Nep}$ = 3.39), 2014 FC$_{69}$  ($q=40.4$ AU, $P / P_{\rm Nep}$ = 3.78), and 2013 JD$_{64}$ ($q=42.6$ AU, $P / P_{\rm Nep}$ = 3.72). Because these objects typically attain their large perihelia through Gyrs of evolution, they are unlikely to be strongly tied to Neptune's early migration history.

\subsection{Low-Perihelion Counterparts}

Although we have focused on the high-perihelion population of TNOs throughout this work, there is of course a much larger population of objects with $q<40$ AU in every simulation, as illustrated in Figure 2. In fact, even many of the non-fossilized objects that finish our simulations with $q>40$ AU periodically move back and forth across the $q=40$ AU boundary as the Kozai mechanism lifts and lowers their perihelia. Thus, some of the known resonant TNOs with $q<40$ AU may actually be a segment of the high-perihelion population that is currently experiencing a low-$q$ Kozai phase. We can estimate the size of the low-$q$ population that may be dynamically linked with the high-perihelion population by studying the fraction of time that resonant particles spend at $q>40$ AU vs $q<40$ AU. To do this, we study particles that first attain $q>40$ AU between $t=500$ Myrs and $t=3000$ Myrs. This eliminates the fossilized population (which will spend all of their subsequent time at high-perihelion) and also ensures that particles spent at least 1 Gyr near or in a resonance. When we look at this sample of particles, we find that these particles spend about 2/3 of their time at $q>40$ AU and 1/3 of their time at $q<40$ AU, regardless of the simulation. Consequently, for every two resonant high-perihelion objects we expect there to be another resonant object in a low-$q$ Kozai phase. 

In addition, there are also many resonant objects that have lower inclinations that are unaffected by the Kozai mechanism and continually stay at $q<40$ AU. Here we estimate the total fraction of the Kuiper belt's mass trapped in well-populated resonances both at low-perihelion and at high-perihelion. To do this, we count the number particles that have final orbital period ratios that are no higher than 0.05 above the resonance orbital period ratio and no less than 0.1 below the resonance period ratio. (This asymmetry is meant to capture some of the resonance trail.) We do this for the 7:3, 5:2, 8:3, 3:1, 7:2, 4:1, 5:1, and 6:1 resonances -- our most populated resonances beyond the 2:1 MMR. Because we do not actually search for resonant angle libration in each particle's history, these should be taken as rough estimates since some non-resonant particles are inevitably included.

The results of this procedure are shown in Figure 13. Here we see several notable features. First, we find that the 7:3 and 5:2 resonances host the largest numbers of particles in every simulation. In spite of this, they never host the largest numbers of high-perihelion objects. Instead, the 3:1 resonance always contains the largest number of high-perihelion objects. Nevertheless, except in our GS simulation, most of the 3:1 particles are typically found with perihelia inside 40 AU. We also notice that that the 8:3 MMR typically contains comparable numbers of particles to the 3:1 but very few are found at high-perihelion. In fact, high-perihelion 8:3 objects are completely absent in our smooth migration simulations. Although our particle numbers are limited, these results and the very existence of 2004 XR$_{190}$ suggests that Neptune's migration had some element of graininess. Finally, we note that beyond the 3:1 there is always a steady fall off in particle number. 

\begin{figure*}
\centering
\includegraphics[scale=0.44]{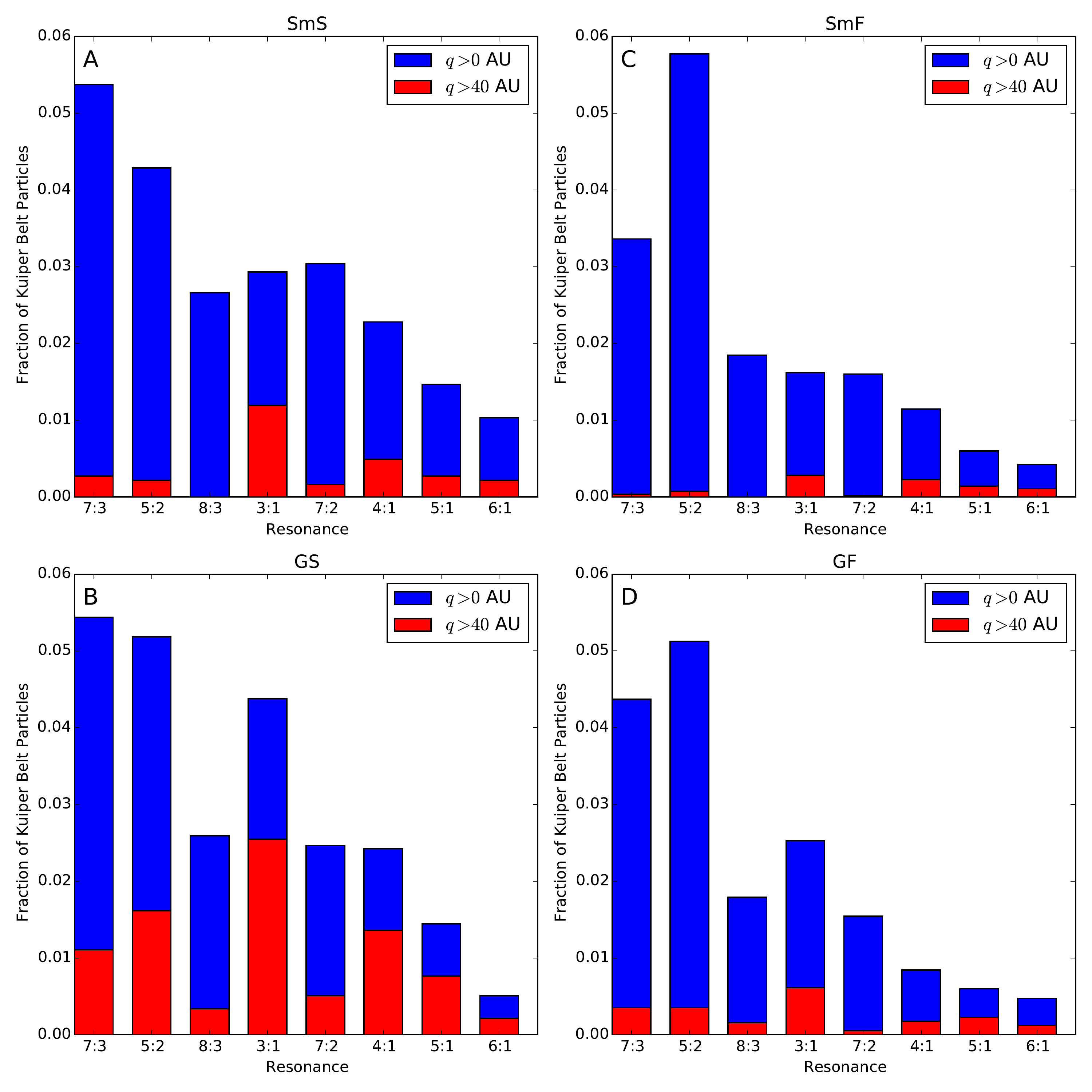}
\caption{Histograms of the fraction of surviving Kuiper belt particles found near several different resonances at the end of our simulations Blue shows the fraction of particles with semimajor axes near the resonance location. Red shows the fraction of particles with semimajor axes near the resonance location {\it and} with $q>40$ AU. Results from our SmS, SmF, GS, and GF simulations are shown in panels A, B, C, and D, respectively.}
\end{figure*}

\section{Discussion and Conclusions}

The high-perihelion population of the Kuiper belt that is resonant or near-resonant with Neptune is a burgeoning subpopulation of detected TNOs that is very well-suited for constraining the migration of Neptune. As Neptune migrates, objects can be captured into these resonances and have their perihelia raised by Kozai cycles. On average we find that objects actively undergoing such Kozai cycles spend $\sim$2/3 of their time with $q>40$ AU and $\sim$1/3 with $q<40$ AU. Because the Kozai mechanism increases orbital inclination as it cycles objects to high perihelia, we predict that the median inclination of resonant TNOs with $q>40$ AU is 34$^{\circ}$, and 99\% of these high-perihelion objects have inclinations over 20$^{\circ}$. As Neptune continues to migrate, these objects can fall out of resonance during the high-perihelion ($q\gtrsim40$ AU) phases of their Kozai cycles, becoming frozen in $a$, $q$, and $i$. This produces a trail of high-perihelion objects that are slightly Sunward of the modern locations of the resonances. However, this trailing population can only exist if Neptune migrates slowly enough. Otherwise, Neptune finishes migrating before resonance capture and Kozai-lifting can actually take place. Hence, the extent of the resonance-trailing population depends on whether Neptune's migration time is shorter or longer than the typical timescale for resonance capture and perihelion-lifting via the Kozai mechanism. 

Regardless of the nature and timescale of Neptune's migration, we find that high-perihelion objects associated with the 3:1 MMR are the easiest to detect due to their overall large number and proximity to the Sun. The 4:1 population can approach or even slightly exceed the 3:1, but this population is about 14 AU more distant in semimajor axis. Meanwhile, a grainy migration history for Neptune can greatly enhance the number of objects that fall out of the 7:3, 5:2, and 8:3 resonances, but individually they still contain less objects than the 3:1 MMR, regardless of Neptune's migration smoothness. The abundance and relative detectability of high-perihelion objects near the 3:1 MMR is consistent with the recent detections of \citet{shep16}, who detect 3 new high perihelion objects near the 3:1, and no more than one object associated with any other resonance.

The 3:1 MMR is also unique among the high-perihelion population of TNOs because objects in this resonance tend to be lifted to high-perihelion on a shorter timescale than any other resonance beyond the 2:1. While the 7:3 also begins lifting the perihelia of objects quickly, they are much less likely to fall out of resonance, unless Neptune's migration is grainy. Beyond the 4:1 MMR, the typical timescale for raising TNO perihelia approaches 1 Gyr, and the generation of these populations is not strongly influenced by Neptune's early migration. They represent a more contemporary set of TNO orbits and should show few trailing objects. This population is dominated by the N:1 resonances, whose individual populations fall off as $N^{-1.4}$.

Because the 3:1 MMR is quickly populated with a large number of objects that can subsequently fall out of resonance if Neptune migrates slowly, we highlight the orbital distribution of high-perihelion objects near this particular resonance as a way to constrain Neptune's migration timescale. We model the formation of the high-perihelion Kuiper belt population using migration scenarios and timescales previously found to be consistent with the rest of the Kuiper belt \citep{nes15a, nes16}. We find that when Neptune reaches its modern orbit within $\sim$100 Myrs, more than 90\% of high-perihelion objects near the 3:1 have semimajor axes that are consistent with orbits than can still be resonantly active with Neptune. Meanwhile, if we just slow Neptune's migration by a factor of 3 so that it takes 300 Myrs to reach its modern location, $\sim$50\% of the high-perihelion population sits $\sim$1 AU closer to the Sun than then modern MMR center. Thus, the distribution of tail of high-perihelion orbits near the 3:1 MMR is a sensitive diagnostic for Neptune's migration timescale.

Currently, there are only 5 high-perihelion objects known near the 3:1 resonance, which is not enough to rule out any of the migration scenarios we explore here with a high degree of confidence. Nevertheless, the fact that at least one of these objects is fossilized (2015 KH$_{162}$) hints at the presence of a significant trailing population for the 3:1 MMR, which suggests that our slow migration simulations may be more consistent with the actual solar system. In addition, without an element of grainy migration, it is more difficult for objects to fall out of the 7:3, 5:2, and 8:3 MMRs. Out of all the resonances we study here, the 7:3, 5:2, and 8:3 MMRs offer the best possibilities to constrain the smoothness of Neptune's migration. Although we cannot use the current sample of high-perihelion objects to make any definitive statements on the smoothness of Neptune's migration, the existence of 2004 XR$_{190}$ near but closer than the modern 8:3 location supports the conclusion of \citet{nes16} that Neptune's migration was grainy.

Because the distribution of high-perihelion objects is so sensitive to relatively minor changes in Neptune's migration timescale, only a modest increase in the sample size of this population may be able to differentiate a 100-Myr from a 300-Myr total migration time for Neptune, or a smooth vs grainy migration style. Furthermore, our simulations already suggest that one known object, 2015 KH$_{162}$, belongs to the fossilized resonance trail of the 3:1. An added benefit of focusing on a single resonant population such as that near the 3:1 MMR is that all of these bodies have a very small range in semimajor axes, so the detected distribution of orbital semimajor axes are less affected by observational biases that favor the detection of closer objects over further ones. In the next several years, ongoing and upcoming surveys should provide a high-perihelion 3:1 population large enough to robustly constrain Neptune's migration \citep{shep16, weryk16, ivez08}.

\section{Acknowledgements}

NAK acknowledges support from a junior faculty fellowship from the University of Oklahoma College of Arts \& Sciences. SSS acknowledges support from the Carnegie Institution for Science. We are grateful for the open discussion and cooperation with David Nesvorny, as we became aware of the similar but independent, simultaneous work of \citet{nes16b} as we neared completion of this manuscript.

\bibliographystyle{apj}
\bibliography{KZMMR}

\end{document}